\documentclass[amsmath,amssymb,aps,notitlepage,twocolumn,showpacs,10pt,superscriptaddress]{revtex4-1}

\usepackage{graphicx}
\usepackage{hyperref}

\newcommand{\be}{\begin{equation}}
\newcommand{\ee}{\end{equation}}
\newcommand{\bea}{\begin{eqnarray}}
\newcommand{\eea}{\end{eqnarray}}

\begin{document}
\title{New constraints on axion-mediated P,T-violating
interaction from electric dipole moments of diamagnetic atoms}

\author{V. A. Dzuba}
\affiliation{School of Physics, University of New South Wales, Sydney, New South Wales 2052, Australia}
\author{V. V. Flambaum}
\affiliation{School of Physics, University of New South Wales, Sydney, New South Wales 2052, Australia}
\author{I. B. Samsonov}
\affiliation{School of Physics, University of New South Wales, Sydney, New South Wales 2052, Australia}
\affiliation{Bogoliubov Laboratory of Theoretical Physics, JINR, Dubna, Moscow region 141980, Russia}
\author{Y. V. Stadnik}
\affiliation{Helmholtz Institute Mainz, Johannes Gutenberg University, 55099 Mainz, Germany}

\begin{abstract}
The exchange of an axion-like particle between atomic electrons
and the nucleus may induce electric dipole moments (EDMs) of atoms
and molecules. This interaction is described by a parity- and
time-reversal-invariance-violating potential which depends on the
product of a scalar $g^s$ and a pseudoscalar $g^p$ coupling
constant. We consider the interaction with the specific
combination of these constants, $g_e^s g_N^p$, which gives
significant contributions to the EDMs of diamagnetic atoms. In
this paper, we calculate these contributions to the EDMs of
$^{199}$Hg, $^{129}$Xe, $^{211}$Rn and $^{225}$Ra for a wide range
of axion masses. Comparing these results with recent experimental
EDM measurements, we place new constraints on $g_e^s g_N^p$. The
most stringent atomic EDM limits come from $^{199}$Hg and improve
on existing laboratory limits from other experiments for axion
masses exceeding $10^{-2}$ eV.
\end{abstract}

\maketitle

\section{Introduction}

In field theory, the interaction of the axion field $a$ with
fermions $\psi$ may be described by the Lagrangian density
\be
{\cal L}_{\rm int} = a \sum_\psi \bar\psi(g_\psi^s + i g_\psi^p
\gamma_5)\psi\,,
\label{axion-interaction}
\ee
where $g_\psi^s$ and $g_\psi^p$ are model-dependent coupling
constants and $\gamma_5 = -i \gamma_0 \gamma_1 \gamma_2 \gamma_3$
in the notation of \cite{Khriplovich-book} for Dirac matrices.
This Lagrangian appears naturally in the case of the canonical axion,
which solves the strong CP problem of quantum chromodynamics
\cite{Peccei:1977hh,Weinberg:1977ma,Wilczek:1977pj,Kim:1979if,Shifman:1979if,Zhitnitsky:1980he,Dine:1981rt}.
In Eq.\ (\ref{axion-interaction}), we assume, however, a
generic axion-like particle, which couples to different fermions
with independent constants $g_\psi^s$ and
$g_\psi^p$. Consistency with various experimental data imposes very severe
constraints on different combinations of such couplings, see, e.g., Ref.\ \cite{Safronova:2017xyt} for
a review. Since
these interactions are extremely weak, the axion can naturally be
considered a candidate for dark matter \cite{Preskill:1982cy,Abbott:1982af,Dine:1982ah}.

%%Igor
In atomic phenomena, the interaction (\ref{axion-interaction})
implies the exchange of an axion between the atomic electrons
and the nucleus described by the P,T-violating potential
\be
V(r) = i \frac{g^p g^s }{4\pi}\frac{e^{-m_a r}}{r}
\gamma^0\gamma_5\,,
\label{V0}
\ee
where $m_a$ is the axion mass. In Ref.\ \cite{Stadnik:2017hpa}, it
was shown that this potential induces anomalous contributions to
EDMs in atoms and molecules due to mixing of atomic states of
opposite parity. The comparison of these EDMs with the corresponding
experimentally observed values imposes strong constraints on the
coupling constants $g^s$ and $g^p$ of the interaction
(\ref{axion-interaction}).

Ref.\ \cite{Stadnik:2017hpa}
considered the case when the pseudoscalar interaction constant $g^p\equiv g_e^p$
is attributed to the electron, while the scalar interaction constant $g^s\equiv g_N^s$
corresponds to either another electron or a nucleon.
% Igor
In the latter case, the potential (\ref{V0}) reduces (in the non-relativistic limit) to
\be
V(r) = -\frac{g_e^p g_N^s}{8\pi m_e }{\boldsymbol \Sigma}\cdot
{\boldsymbol\nabla}\left(
\frac{e^{-m_a r}}{r}
\right)\,,
\label{V12old}
\ee
where $m_e$ is the electron mass and ${\boldsymbol \Sigma}=\left(
\begin{array}{cc}{\boldsymbol\sigma} & 0 \\ 0 & {\boldsymbol\sigma}  \end{array}
\right)$ is the Dirac spin matrix vector acting on the electron wavefunctions.
Analysis of contributions to atomic EDMs due to the potential
(\ref{V12old}) allowed the authors to place constraints on the product of
coupling constants $g_e^p g_N^s$. The constraints derived in Ref.\ \cite{Stadnik:2017hpa}
gave a significant improvement over previous laboratory limits on these interaction constants
for certain axion masses \cite{19,21,24,29,37,38,39}.

In this paper, we consider the opposite
case, namely when the constant $g^s\equiv g_e^s$ corresponds to the
interaction of the axion with an electron, while $g^p\equiv g_N^p$ corresponds to
the interaction with a nucleon. In this case, the potential
(\ref{V0}) reduces to the following form
\be
V(r) = -\frac{g_e^s g_N^p}{8\pi m_N} \boldsymbol{\sigma}_N\cdot\boldsymbol{\nabla}
\left(
\frac{e^{-m_a r}}{r}
\right)\gamma^0\,,
\label{V12}
\ee
where $m_N$ and $\boldsymbol{\sigma}_N$ are the nucleon mass and its spin unit
vector, respectively. The Dirac matrix $\gamma^0$ corresponds to the atomic
electrons. The potential (\ref{V12}) will allow us to place new
constraints on the combination of coupling constants $g_e^s g_N^p$,
which is independent from the case considered in
\cite{Stadnik:2017hpa}.

Although the potentials (\ref{V12old}) and (\ref{V12}) look
similar, they manifest themselves differently in atomic
phenomena. The potential (\ref{V12old}) describes the interaction
of electron's spin with the nuclear density. Thus,
this interaction may give significant contributions to the atomic
EDMs of paramagnetic atoms with open electron shells. On the other hand, the potential
(\ref{V12}) is responsible for the interaction of the nuclear spin
with the electron density, which may contribute significantly to the
atomic EDMs of diamagnetic atoms with closed electron shells,
but with non-zero nuclear spins.
%%%%%%%%%%%%%%%%%%%%%%%%%%%%%%%%%%%%%%%%%%%%%
Therefore, we perform numerical calculations of the corresponding EDMs for atomic
$^{129}$Xe, $^{199}$Hg, $^{211}$Rn and $^{225}$Ra to interpret existing experimental
data. We find that the most stringent constraint arises from
the recent EDM measurement in $^{199}$Hg \cite{Graner:2016ses}:
\begin{equation}
d(^{199}{\rm Hg})=(2.20 \pm 2.75_{\rm stat} \pm 1.48_{\rm syst})
\times 10^{-30} e\ {\rm cm}\,.\label{199HgEDM}
\end{equation}
This allows us to place new bounds on the combination of coupling
constants $g_e^s g_N^p $ for a wide range of axion
masses. Measurements of EDMs in the other diamagnetic atoms
\cite{NewXe,Xe,Bishof:2016uqx} give less stringent constraints.

We note that in the limit of a large axion mass, the potential
(\ref{V12}) reduces to the following contact interaction
\be
\lim_{m_a\to\infty} m_a^2 V(r)  = -\frac{g_e^s g_N^p}{2m_N}
\boldsymbol{\sigma}_N\cdot\boldsymbol{\nabla}[ \delta^3({\bf r})]\gamma^0\,.
\label{V12-lim}
\ee
In field theory, this potential corresponds to the
parity- and time-reversal-invariance-violating four-fermion interaction
Lagrangian density
\be
{\cal L} = -\frac{G_F}{\sqrt2}  C_{PS} \bar N i\gamma_5 N \bar e e\,,
\ee
where $N$ and $e$ denote the nucleon and
electron fields, respectively; $G_F$ is the Fermi constant and
$C_{PS}=-\sqrt2 g_e^s g_N^p/(G_F m_a^2)$.
The contributions to the EDMs of diamagnetic atoms due to this
operator were studied in previous works
\cite{Flambaum:1985gx,Dzuba:2009kn}. In the next section,
we extend these earlier calculations to the potential (\ref{V12}), which
is defined for an arbitrary axion mass.

\section{Calculations and results}

The potential (\ref{V12}) describes the interaction of a
non-polarized electron with a polarized nucleon with spin $\boldsymbol{\sigma}_N$.
To apply this potential to the electron-nucleus interaction in an atom, we have to average
it over the atomic nucleus,
\be
\bar V= -\frac{g_e^s g_N^p}{8\pi m_N} \langle \boldsymbol{\sigma}_N\rangle
\cdot
\left\langle
\boldsymbol{\nabla}\left(
\frac{e^{-m_a r}}{r}\right)\right\rangle \gamma^0\,.
\label{4}
\ee
Here, the quantity $\langle\boldsymbol{\sigma}_N\rangle$ is proportional to the total
angular momentum of the nucleus $\bf I$
\be
\langle\boldsymbol{\sigma}_N\rangle = \kappa\,{\bf
I}/I\,.
\label{5}
\ee
For spherically symmetric non-excited nuclei $^{129}$Xe and
$^{199}$Hg the coefficient $\kappa$ can be computed within the Schmidt
(single-particle approximation) model \cite{Dzuba:2009kn,Stadnik:2014xja}
(see also Refs.\ \cite{Yoshinaga1,Yoshinaga2,Yoshinaga3},
which employ more sophisticated nuclear models).
For the deformed nuclei $^{211}$Rn and $^{225}$Ra it is
appropriate to use the Nilsson nuclear model (see, e.g., \cite{book2}). We collect the
resulting values of the coefficient $\kappa$ in Table \ref{Tab0}.
We point out that the spins of
these nuclei are predominantly due to the spin of the
unpaired valence neutron, irrespective of the nuclear model used. Thus,
experimental measurements of EDMs of these atoms are
mainly sensitive to the parameter $g_n^p$, which
corresponds to the interaction of an axion with a neutron.
\begin{table}[htb]
\caption{Values of the coefficient $\kappa$ in Eq.\ (\ref5) for different nuclei.
For spherically symmetric nuclei $^{129}$Xe and $^{199}$Hg this
coefficient is found using the Schmidt nuclear model while for the
deformed nuclei $^{211}$Rn and $^{225}$Ra the Nilsson deformed oscillator model is applied.}
\begin{tabular}{c| c |c | c|c }\hline\hline
 & $^{129}$Xe & $^{199}$Hg & $^{211}$Rn & $^{225}$Ra \\\hline
$\kappa$ & $+1$ & $-\frac13$ & $-\frac13$ & $\frac13$\\\hline\hline
\end{tabular}
\label{Tab0}
\end{table}

The potential (\ref4) involves the Yukawa-type
interaction which should be averaged over the nuclear density
$\rho({\bf R})$,
\be
\left\langle
\frac{e^{-m_a r}}{r}\right\rangle =
\int d^3{\bf R}
\frac{e^{-m_a |{\bf r}-{\bf R}|}}{|{\bf r}-{\bf R}|} \rho({\bf
R})\,.
\label{6}
\ee
Note that, according to the experimental data discussed in  Ref.~\cite{Khriplovich-book},
the shape of the nuclear spin density is very close to the shape of the nuclear charge density.
The nuclear density is well described by the Fermi function
\be
\rho({\bf R}) = \frac{\rho_0}{1+\exp\left(\frac{R-R_0}{a}\right)}\,,
\label{rho1}
\ee
where $R_0$ and $a$ are nucleus-dependent parameters and $\rho_0$ is
normalized according to
$\int \rho({\bf R})d^3{\bf R} =1$. The values of these parameters for various isotopes are
tabulated, e.g., in \cite{Fricke2004}.
%%%%%%%%%%%%%%%%%%%%%%%%%%%%%%%%%%%%%%%%%%%%
The potential (\ref4), averaged over the nuclear density (\ref{rho1}), takes the following radial form
\bea
&& V_r(r)= -\frac{\kappa G_F C_{PS}m_a^3}{2\sqrt2 m_N} \left[ -\frac{e^{-m_ar}}{m_a^2r^2} (1+m_ar) \right.  \nonumber \\
&& \times\int_0^r r'\rho(r')\sinh(m_ar')dr' +  \frac{1}{m_a^2r^2}\left[ m_ar\cosh(m_ar) \right. \nonumber \\
&&\left. \left. - \sinh(m_ar)\right] \int_r^{\infty} r'\rho(r')e^{-m_ar'}dr' \right].
\label{eq:V}
\eea
With the single-electron wave function of the form
\be
 \psi({\bf r}) = \frac{1}{r} \left(\begin{array}{c} f(r) \Omega_{jlm} \\ ig(r) \Omega_{j \tilde l m} \end{array} \right),
 \label{eq:wf}
\ee
where $ \Omega_{jlm}$ is the spherical spinor,
the single-electron matrix element of the operator (\ref4) is given by
\bea
&&\langle i|| \bar V || j \rangle = \langle \kappa_i || C^{(1)} || \kappa_j\rangle \times \label{eq:me} \\
&&\int V_r(r) \left[ f_i(r)f_j(r) - g_i(r)g_j(r)\right] dr, \nonumber
\eea
where $\kappa=(-1)^{j+1/2-l}(j+1/2)$ is the angular quantum number which determines the angular
momentum $l$ and total angular momentum $j$; the potential $V(r)$ is given by (\ref{eq:V}), and $ C^{(1)}$ is the normalised spherical
function of first rank.
%%%%%%%%%%%%%%%%%%%%%%%%%%%%%%%%%%%%%%%%%%%%

The interaction potential (\ref{4}) induces an atomic EDM
which in the leading order in perturbation theory reads
\be
{\bf d} = 2 \sum_M \frac{\langle0|{\bar V}|M\rangle \langle M|-e{\bf r}|0\rangle}{E_0 -
E_M}\,,
\label{EDM}
\ee
where $e$ is the electron charge and the summation is over the complete set of excited states
$|M\rangle$ with energies $E_M$.
%%%%%%%%%%%%%%%%%%%%%%%%%%%%
Note that if the operator $\bar V$ is replaced in (\ref{EDM}) by the electric dipole operator $-e{\bf r}$, then
this expression gives the scalar static polarizability of the atom. This can be used to check the accuracy of
the calculations.

Eq.~(\ref{EDM}) is exact if $|0\rangle$ and $|M\rangle$ are exact many-electron wave functions of the whole
atom. In practice, we need to reduce the calculations to single-electron matrix elements while including many-body
effects as corrections to the wave function or to the operator. In the present work, we use the self-consistent
%Victor minor corrections  below
relativistic Dirac-Hartree-Fock method in the external field, also known as the random-phase approximation (RPA). This way we
include the electron core polarization corrections to the operator of the external field.
 All electron states from the closed shells are treated as the core states producing the Hartree-Fock potential. In the closed-shell atoms which we consider in this paper, all electrons are included in the electron core.

The RPA equations can be written in terms of the corrections to all electron wave functions induced by the external field
\be
(\hat H^{\rm HF} - \epsilon_c) \delta \psi_c = - (\hat F + \delta V^F) \psi_c.
\label{eq:RPA}
\ee
Here $\hat H^{\rm HF}$ is the relativistic Hartree-Fock operator, $\psi_c$ is the single-electron wave function
for a  state $c$, $\hat F$ is the operator of the external field (either electric dipole ${\bf D}=-e{\bf r}$ or
parity-violating operator (\ref4)), $\delta V^F$ is the correction to the self-consistent Hartree-Fock
potential (including the exchange interaction) due to the change of all electron  states imposed by the external field. Equations (\ref{eq:RPA}) are solved
self-consistently for all states in the core. After that, the EDM of a closed-shell atom in the RPA approximation
is given by
\be
{\bf d} = \frac{2}{3} \sum_c \langle \psi_c || \bar V || \delta \psi_c^d \rangle \equiv
\frac{2}{3} \sum_c \langle \psi_c || {\bf D} || \delta \psi_c^{\bar V} \rangle. \label{RPAEDM}
\ee
Here $\delta \psi_c^d$ comes from solving the RPA equations (\ref{eq:RPA}) with the electric dipole operator
($\hat F = {\bf D}$), while $\delta \psi_c^{\bar V}$ comes from solving the RPA equations  with the operator $\bar V$.
The two equations in (\ref{RPAEDM}) are equivalent and comparing the results can be used to check
computer codes. Note that it is sufficient to solve the RPA equations only once with either of the two operators.
As in Eq.~(\ref{EDM}), replacing the $\bar V$ operator in (\ref{RPAEDM}) by the electric dipole operator gives the
scalar dipole polarizability of the atom in the RPA approximation.
It is known that the RPA approximation
give very good accuracy for the polarizabilities of noble-gas atoms (see, e.g. \cite{Dzuba16}). The difference
with experiment is at the level of 1 -- 4\% for Kr and Xe. On the other hand, the difference is larger for atoms like Hg and Ra.
These atoms have $6s^2$ and $7s^2$ outermost subshells. Therefore, we can still treat them as closed-shell
systems and use (\ref{eq:RPA}) and (\ref{RPAEDM}) for the calculations. The difference with experiment is
 7\% for the polarizability of Ra and 25\% for the polarizability of Hg~\cite{Dzuba16}. This difference between
calculated and measured polarizabilities is mostly due to many-body effects beyond the core polarization.
This is the dominating source of uncertainty in the present atomic calculations.
The contribution of these correlations to the EDM of Hg, Yb, Ra and other atoms was studied
in detail for a range of singular operators in our earlier works \cite{Dzuba:2009kn,Kozlov}.
It was found that this contribution does not exceed 15\%.
Other factors, like numerical accuracy in
calculating the potential (\ref{eq:V}) or solving the RPA equations (\ref{eq:RPA}), or uncertainty in the nuclear
parameters in (\ref{rho1}), among others, give much smaller contributions to the uncertainties.
Thus we conclude that our accuracy for
the atomic EDM calculations is within 10\% for Xe and Rn and 30\% for Hg and Ra.
%%%%%%%%%%%%%%%%%%%%%%%

The results of calculations are summarized in Table~\ref{Tab1}.
In this table, the results for an infinite axion mass are
taken from \cite{Dzuba:2009kn}, where the EDMs induced by
the operator (\ref{V12-lim}) were calculated.

\begin{table}[htb]
\caption{Summary of relativistic Hartree-Fock-Dirac calculations of atomic EDMs induced
by the interaction (\ref4) for various axion masses. These values are given in
the units $C_{PS} \cdot \kappa\cdot e
\cdot {\rm cm}$, where $C_{PS}=-\sqrt2g_e^s g_N^p/(G_F m_a^2)$.
These calculations take into account the effects of all atomic electrons.
For axion masses $m_a\ll 1$ keV, the interaction (\ref{4}) becomes long-range and the
induced atomic EDMs become independent of $m_a$. The estimated accuracy is within 10\% for
Xe and Rn and 30\% for Hg and Ra.}
\begin{tabular}{c| c |c| c |c}\hline\hline
$m_a$  &  $^{129}$Xe & $^{199}$Hg  & $^{211}$Rn & $^{225}$Ra
\\
(eV) &&&& \\\hline
$\infty$ & $1.6\times 10^{-23}$ & $-1.8\times 10^{-22}$ & $2.1\times 10^{-22}$ & $-6.4\times 10^{-22}$ \\
$10^8$   & $1.4\times 10^{-23}$ & $-1.8\times 10^{-22}$ & $1.7\times 10^{-22}$ & $-5.2\times 10^{-22}$\\
$10^7$   & $3.6\times 10^{-24}$ & $-3.7\times 10^{-23}$ & $3.5\times 10^{-23}$ & $-1.0\times 10^{-22}$\\
$10^6$   & $5.4\times 10^{-25}$ & $-2.4\times 10^{-24}$ & $2.1\times 10^{-24}$ & $-5.4\times 10^{-24}$\\
$10^5$   & $8.9\times 10^{-27}$ & $-2.7\times 10^{-26}$ & $1.7\times 10^{-26}$ & $-5.5\times 10^{-26}$\\
$10^4$   & $4.2\times 10^{-29}$ & $-2.0\times 10^{-28}$ & $1.5\times 10^{-28}$ & $-4.5\times 10^{-28}$\\
$10^3$   & $1.1\times 10^{-30}$ & $-1.0\times 10^{-30}$ & $2.1\times 10^{-30}$ & $-3.7\times 10^{-30}$\\
$10^2$   & $1.2\times 10^{-32}$ & $-7.8\times 10^{-33}$ & $2.3\times 10^{-32}$ & $-3.1\times 10^{-32}$\\
$10$     & $1.2\times 10^{-34}$ & $-7.8\times 10^{-35}$ & $2.3\times 10^{-34}$ & $-3.1\times
10^{-34}$\\\hline\hline
\end{tabular}
\label{Tab1}
\end{table}

\begin{table}[htb]
\caption{Asymptotic values of EDMs of atoms for low and high axion mass.
The values in this table originate
from the corresponding values in Table \ref{Tab1} after substitution of
the value of the Fermi coupling constant $G_F\approx 1.167\times 10^{-5}$ GeV$^{-2}$ and
the coefficient $\kappa$ from Table \ref{Tab0}.}
\begin{tabular}{c| c |c}\hline\hline
$|d|,\ e\cdot\mbox{cm}$ & $m_a\lesssim 10^3$ eV & $m_a \gtrsim 10^8$ eV \\\hline
$^{129}$Xe & $1.5\times 10^{-13}g_e^s g_N^p$ & $1.7\, g_e^s g_N^p\left(
  \frac{\rm eV}{m_a} \right)^2$ \\\hline
$^{199}$Hg & $3.2\times 10^{-14}g_e^s g_N^p$ & $7.3\, g_e^s g_N^p\left(
  \frac{\rm eV}{m_a} \right)^2$ \\\hline
$^{211}$Rn & $9.3\times 10^{-14}g_e^s g_N^p$ & $8.5\, g_e^s g_N^p\left(
  \frac{\rm eV}{m_a} \right)^2$ \\\hline
$^{225}$Ra & $1.3\times 10^{-13}g_e^s g_N^p$ & $25\, g_e^s g_N^p\left(
  \frac{\rm eV}{m_a} \right)^2$ \\\hline\hline
\end{tabular}
\label{Tab2}
\end{table}

\begin{figure}[htb]
\includegraphics[width=8cm]{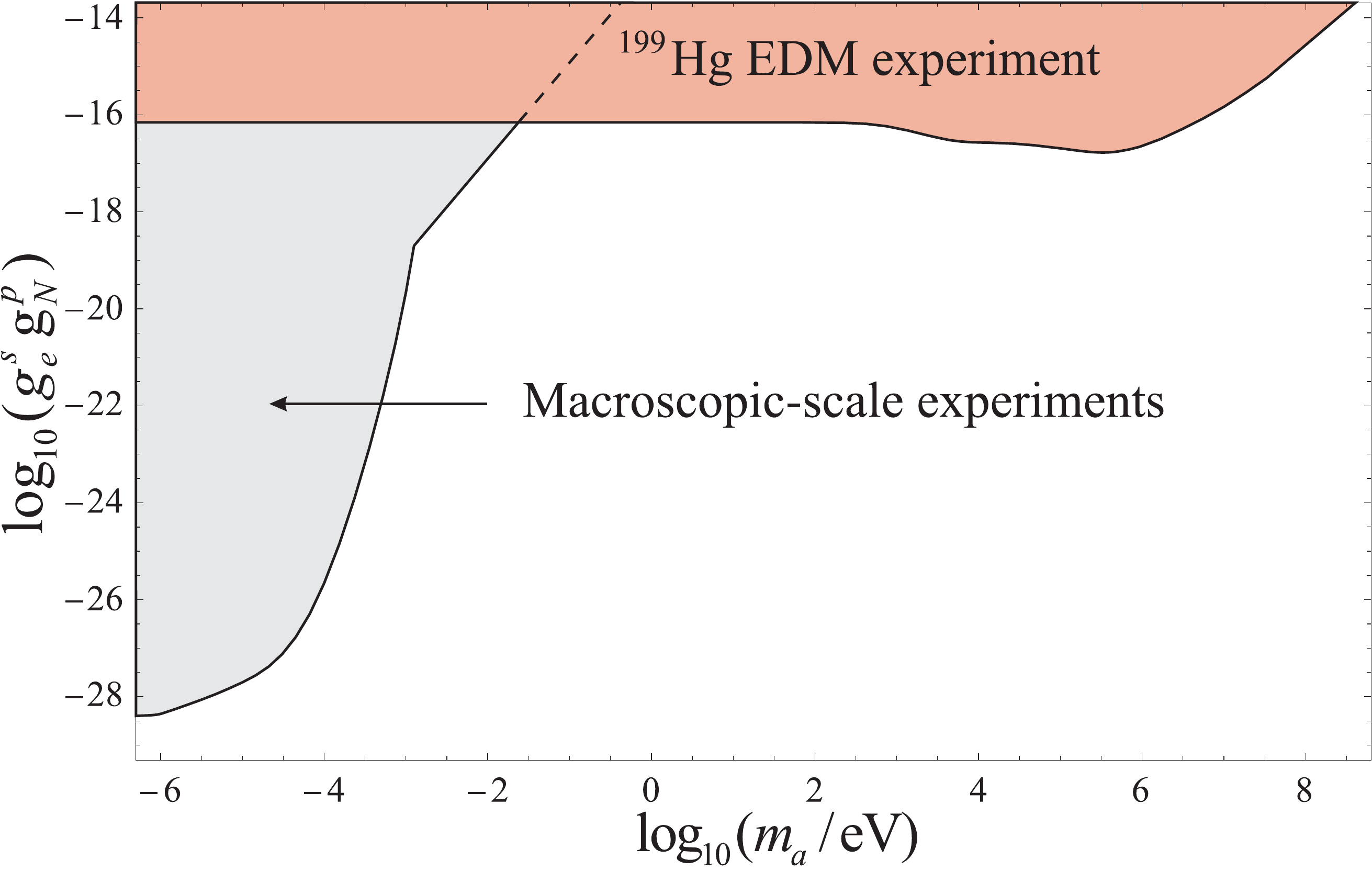}
\caption{Laboratory constraints on the parity- and time-reversal-invariance-violating
scalar-pseudoscalar electron-nucleon interaction mediated
by an axion of mass $m_a$. The pink exclusion region is the result of this work.
The gray exclusion region summarizes the combined results
which were derived from the earlier macroscopic-scale experiments
\cite{SEREBROV2009423,PhysRevLett.105.170401,PhysRevLett.111.100801,PhysRevLett.111.102001,Afach:2014bir}
with graphical accuracy.
}
\label{Fig1}
\end{figure}

\section{Discussion}

In Table \ref{Tab1}, we present the results of our computations of
the EDMs in four diamagnetic atoms ($^{129}$Xe, $^{199}$Hg,  $^{211}$Rn and
$^{225}$Ra) induced by the P,T-odd potential (\ref{V12}).
In Table \ref{Tab2}, we also collect the asymptotical values of
EDMs for these atoms at low ($m_a\lesssim 10^3$ eV) and high
($m_a \gtrsim 10^8$ eV) axion masses.
Combining these results with experimental measurements of
EDMs in these atoms imposes constraints on the product of coupling
constants $g_e^s g_N^p$. The most stringent constraint comes
from the $^{199}$Hg EDM experiment \cite{Graner:2016ses} given in Eq.\ (\ref{199HgEDM})
and is shown in Fig.\ \ref{Fig1} by the pink exclusion region.
The pink exclusion region in Fig.\ \ref{Fig1} possesses the
following asymptotics:
\begin{equation}
\begin{array}{ll}
m_a \lesssim 10^3 \mbox{ eV}\quad & |g_e^s g_N^p| < 7\times
10^{-17}\\
m_a \gtrsim 10^8 \mbox{ eV} &
|g_e^s g_N^p| < 3\times 10^{-31}\left(\frac{m_a}{\rm eV}\right)^2\,.
\end{array}
\end{equation}
The latter constraint originates from the results of the paper
\cite{Dzuba:2009kn}, where the atomic EDMs
due to the operator (\ref{V12-lim}) were studied.

It is interesting to compare our results with earlier
constraints from macroscopic-scale experiments
\cite{SEREBROV2009423,PhysRevLett.105.170401,PhysRevLett.111.100801,PhysRevLett.111.102001,Afach:2014bir},
which reported constraints on the coupling parameters
$g_N^s g_n^p$, where $g_n^p$ denotes the axion coupling to a polarized
neutron, while $g_N^s$ denotes the coupling to the nucleons in a non-polarized massive
body. Let $\langle A\rangle$ and $\langle Z\rangle$ be the average atomic mass
and proton numbers in the non-polarized massive body, respectively. Then, in general, the polarized neutron
interacts with a non-polarized atom through the combination of
constants $(g_e^s \langle Z\rangle + g_N^s \langle
A\rangle)g_n^p$. %, where $g^s_N \langle A\rangle \equiv g^s_n (\langle A\rangle-\langle Z\rangle)  + g_p^s \langle Z\rangle $.
%Here $g^s_n$ and $g^s_p$ denote the scalar coupling constants to neutron and proton, respectively.
The constraints on $g_N^s g_n^p$ were obtained in
\cite{SEREBROV2009423,PhysRevLett.105.170401,PhysRevLett.111.100801,PhysRevLett.111.102001,Afach:2014bir}
with the assumption
$\langle A\rangle |g_N^s| \gg \langle Z\rangle |g_e^s|$,
but we can assume the opposite case,
$\langle A\rangle |g_N^s| \ll \langle Z\rangle |g_e^s|$, to find
the constraints on $ g_e^s g_n^p$. Since different experiments deal
with different materials, we make the simple approximation $\langle A \rangle/\langle Z
\rangle \approx 2.2$. This allows us to represent the results of
the earlier works
\cite{Afach:2014bir,PhysRevLett.111.100801,PhysRevLett.111.102001,PhysRevLett.105.170401,SEREBROV2009423}
in the form of the gray exclusion region in Fig.\ \ref{Fig1}.
We conclude that our results give significantly improved laboratory limits
on $g_e^s g_N^p$ for $m_a \gtrsim 10^{-2}$ eV.

We note that there are more stringent indirect bounds from the combination of stellar energy-loss arguments and laboratory searches for spin-independent fifth forces \cite{Raffelt_2012} or from the combination of stellar energy-loss arguments in several different astrophysical systems \cite{Raffelt_1990,Raffelt_2008,Hardy_2017,Essig_2018} for certain axion masses, though astrophysical bounds may be evaded by mechanisms that inhibit the processes of stellar ``cooling'' via axion emission \cite{Masso_2005,Jain_2005,Jaeckel_2006}.
Finally, we mention that limits on the nucleon-nucleon interaction constants
$g_N^s g_{N'}^p$ have been
derived from the consideration of the nuclear Schiff moments
induced by the exchange of a low-mass axion-like particle between
nucleons within a nucleus \cite{Mantry}.

\vspace{3mm}
{\bf Acknowledgments.}
The authors are grateful to Bryce Lackenby for useful comments.
This work is supported by the Australian Research Council Grant No. PD150101405.
V.V.F.\ acknowledges support from the Gutenberg Fellowship and New Zealand Institute for Advanced Studies.
Y.V.S.\ was supported by the Humboldt Research Fellowship.

%\bibliography{Literature}

%

\end{document}